\documentclass[fleqn]{llncs}
\usepackage{pisij}
\usepackage{version}

\title{Indirect Jumps Improve \\ Instruction Sequence Performance}
\author{J.A. Bergstra \and C.A. Middelburg}
\institute{Informatics Institute, Faculty of Science,
           University of Amsterdam, \\
           Science Park~904, 1098~XH Amsterdam, the Netherlands \\
           \email{J.A.Bergstra@uva.nl,C.A.Middelburg@uva.nl}}

\begin{document}

\maketitle

\begin{abstract}
Instruction sequences with direct and indirect jump instructions are as
expressive as instruction sequences with direct jump instructions only.
We show that, in the case where the number of instructions is not
bounded, we are faced with increases of the maximal internal delays of
instruction sequences on execution that are not bounded by a linear
function if we strive for acceptable increases of the lengths of
instruction sequences on elimination of indirect jump instructions.
\begin{classcode}
D.3.3, F.1.1, F.3.3.
\end{classcode}
\end{abstract}

\section{Introduction}
\label{sect-introduction}

Although instruction sequences with direct and indirect jump
instructions are as expressive as instruction sequences with direct
jump instructions only (see~\cite{BM07e}), indirect jump instructions
are widely used to implement certain features of contemporary
high-level programming languages such as the switch statements and
virtual method calls of Java~\cite{GJSB00a} and C\#~\cite{HWG03a}.
Therefore, we consider a further analysis of indirect jump instructions
relevant.

In this paper, we study the effect of eliminating indirect jump
instructions from instruction sequences with direct and indirect jump
instructions on the performance of instruction sequences with respect
to the interaction with their environment on execution.
It is implicit that the elimination of indirect jump
instructions must preserves the behaviour of the instruction sequence
concerned on execution.
We show that, in the case where the number of instructions is not
bounded, there exist instruction sequences with direct and indirect
jump instructions from which elimination of indirect jump instructions
is possible without a super-linear increase of their maximal internal
delay on execution only if their lengths increase super-linearly on
elimination of indirect jump instructions.%
\footnote
{A super-linear increase means an increase that is not bounded by a
 linear function.}

The work presented in this paper belongs to a line of research whose
working hypothesis is that instruction sequence is a central notion of
computer science.
In this line of research, program algebra~\cite{BL02a} is the setting
used for investigating instruction sequences.
The starting-point of program algebra is the perception of a program as
a single-pass instruction sequence, i.e.\ a finite or infinite sequence
of instructions of which each instruction is executed at most once and
can be dropped after it has been executed or jumped over.
This perception is simple, appealing, and links up with practice.

The program notation used in this paper to show that indirect jumps
improve instruction sequence performance is \PGLBij.
This program notation is a minor variant of \PGLCij, a program notation
with indirect jumps instructions introduced in~\cite{BM07e}.
Both program notations are rooted in program algebra, are closer to
existing assembly languages than the notation provided by program
algebra, and have relative jump instructions.
The main difference between them is that \PGLBij\ has an explicit
termination instruction and \PGLCij\ has not.
This difference makes the former program notation more convenient for
the purpose of this paper.

The performance measure used in this paper is the maximal internal delay
of an instruction sequence on execution.
The maximal internal delay of an instruction sequence on execution is the
largest possible delay that can take place between successively executed
instructions whose effects are observable externally.
Another conceivable performance measure is the largest possible sum of
such delays on execution of the instruction sequence.
In this paper, we do not consider the latter performance measure because
it looks to be less adequate to the interactive performance of
instruction sequences.%
\footnote
{Interactive performance means performance with respect to the
 interaction with the environment on execution.}

The work presented in this paper can be looked at in a wider context.
In the literature on computer architecture, hardly anything can be
found that contributes to a sound understanding of the effects of the
presence of common kinds of instructions in the instruction set of a
computer on important points such as instruction sequence size and
instruction sequence performance.
The work presented in this paper can be considered a first step towards
such an understanding.

This paper is organized as follows.
First, we give a survey of the program notation \PGLBij\
(Sect.~\ref{sect-PGLBij}).
Next, we introduce the notion of maximal internal delay of a \PGLBij\
program (Sect.~\ref{sect-delays}).
After that, we present the above-mentioned result concerning the
elimination of indirect jump instructions
(Sect.~\ref{sect-PGLBij-to-PGLB}).
We also relate the work presented in this paper to the point of view
which is the origin of the line of research to which it belongs
(Sect.~\ref{sect-projectionism}).
Finally, we make some concluding remarks
(Sect.~\ref{sect-conclusions}).

\section{\PGLB\ with Indirect Jumps}
\label{sect-PGLBij}

In this section, we give a survey of the program notation \PGLBij.
This program notation is a variant of the program notation \PGLB, which
belongs to a hierarchy of program notations rooted in program algebra
(see~\cite{BL02a}).
\PGLB\ and \PGLBij\ are closer to existing assembly languages than the
notation provided by program algebra.

It is assumed that fixed but arbitrary numbers $\maxr$ and $\maxn$ have
been given, which are considered the number of registers available and
the greatest natural number that can be contained in a register.
Moreover, it is also assumed that fixed but arbitrary finite sets
$\Foci$ of \emph{foci} and $\Meth$ of \emph{methods} have been given.

The set $\BInstr$ of \emph{basic instructions} is
$\set{f.m \where f \in \Foci, m \in \Meth}$.
The view is that the execution environment of a \PGLBij\ program
provides a number of services, that each focus plays the role of a name
of a service, that each method plays the role of a command that a
service can be requested to process, and that the execution of a basic
instruction $f.m$ amounts to making a request to the service named $f$
to process command $m$.
The intuition is that the processing of the command $m$ may modify the
state of the service named $f$ and that the service in question will
produce $\True$ or $\False$ at its completion.

For example, a service may be able to process methods for setting the
content of a Boolean cell to $\True$ or $\False$ and a method for
getting the content of the Boolean cell.
Processing of a setting method may modify the state of the service,
because it may change the content of the Boolean cell with which the
service deals, and simply produces the final content at its completion.
Processing of the getting method does not modify the state of the
service, because it does not changes the content of the Boolean cell
with which the service deals, and produces the content of the Boolean
cell at its completion.

\PGLBij\ has the following primitive instructions:
\begin{iteml}
\item
for each $a \in \BInstr$, a \emph{plain basic instruction} $a$;
\item
for each $a \in \BInstr$, a \emph{positive test instruction} $\ptst{a}$;
\item
for each $a \in \BInstr$, a \emph{negative test instruction} $\ntst{a}$;
\item
for each $l \in \Nat$, a \emph{direct forward jump instruction}
$\fjmp{l}$;
\item
for each $l \in \Nat$, a \emph{direct backward jump instruction}
$\bjmp{l}$;
\item
for each $i \in [1,\maxr]$ and $n \in [1,\maxn]$,
a \emph{register set instruction} $\setr{i}{n}$;
\item
for each $i \in [1,\maxr]$, an \emph{indirect forward jump instruction}
$\ifjmp{i}$;
\item
for each $i \in [1,\maxr]$, an \emph{indirect backward jump instruction}
$\ibjmp{i}$;
\item
a \emph{termination instruction} $\halt$.
\end{iteml}
\PGLBij\ programs are expressions of the form
$u_1 \conc \ldots \conc u_k$,
where $u_1,\ldots,u_k$ are primitive instructions of \PGLBij.
\PGLB\ programs are \PGLBij\ programs in which register set
instructions, indirect forward jump instructions and indirect backward
jump instructions do not occur.

On execution of a \PGLBij\ program, the primitive instructions of
\PGLBij\ have the following effects:
\begin{itemize}
\item
the effect of a positive test instruction $\ptst{a}$ is that basic
instruction $a$ is executed and execution proceeds with the next
primitive instruction if $\True$ is produced and otherwise the next
primitive instruction is skipped and execution proceeds with the
primitive instruction following the skipped one -- if there is no
primitive instruction to proceed with, inaction occurs;
\item
the effect of a negative test instruction $\ntst{a}$ is the same as the
effect of $\ptst{a}$, but with the role of the value produced reversed;
\item
the effect of a plain basic instruction $a$ is the same as the effect of
$\ptst{a}$,\linebreak[2] but execution always proceeds as if $\True$ is
produced;
\item
the effect of a direct forward jump instruction $\fjmp{l}$ is that
execution pro\-ceeds\nolinebreak[2] with the $l$-th next instruction of
the program concerned -- if $l$ equals\linebreak[2] $0$ or there is no
primitive instruction to proceed with, inaction occurs;
\item
the effect of a direct backward jump instruction $\bjmp{l}$ is that
execution proceeds with the $l$-th previous instruction of the program
concerned -- if $l$ equals $0$ or there is no primitive instruction to
proceed with, inaction occurs;
\item
the effect of a register set instruction $\setr{i}{n}$ is that the
content of register $i$ is set to $n$ and execution proceeds with the
next primitive instruction -- if there is no primitive instruction to
proceed with, inaction occurs;
\item
the effect of an indirect forward jump instruction $\ifjmp{i}$ is the
same as the\linebreak[2] effect of $\fjmp{l}$, where $l$ is the content
of register $i$;
\item
the effect of an indirect backward jump instruction $\ibjmp{i}$ is the
same as the effect of $\bjmp{l}$, where $l$ is the content of register
$i$;
\item
the effect of the termination instruction $\halt$ is that execution
terminates.
\end{itemize}
If execution proceeds unbroken and forever with no other primitive
instructions than jump instructions and register set instructions, this
is identified with inaction.

\PGLBij\ is a minor variant of \PGLCij, a program notation with indirect
jumps instructions introduced in~\cite{BM07e}.
The differences between \PGLBij\ and \PGLCij\ are the following:
\begin{itemize}
\item
in those cases where inaction occurs on execution of \PGLBij\ programs
because there is no primitive instructions to proceed with, termination
takes place on execution of \PGLCij\ programs;
\item
the termination instruction $\halt$ is not available in \PGLCij.
\end{itemize}
The meaning of \PGLCij\ programs is formally described in~\cite{BM07e}
by means of a mapping of \PGLCij\ programs to closed terms of program
algebra.
In that way, the behaviour of \PGLCij\ programs on execution is
described indirectly: the behaviour of the programs denoted by closed
terms of program algebra on execution is formally described in several
papers, including~\cite{BM07e}, using basic thread algebra~\cite{BL02a}.%
\footnote
{In several early papers, basic thread algebra is presented under the
name basic polarized process algebra.}
Because \PGLBij\ is a minor variant of \PGLCij, we refrain from
describing the behaviour of \PGLBij\ programs on execution formally in
the current paper.

\section{Internal Delays of \PGLBij\ Programs}
\label{sect-delays}

In this section, we will define the notion of maximal internal delay of
a \PGLBij\ program.

It is assumed that a fixed but arbitrary set $\Aux \subset \BInstr$ of
\emph{auxiliary basic instructions} has been given.
The view is that, in common with the effect of jump instructions and
register set instructions, the effect of auxiliary basic instructions
is wholly unobservable externally, but contributes to the realization
of externally observable behaviour.
Typical examples of auxiliary basic instructions are basic instructions
for storing and fetching data of a temporary character.
Typical examples of non-auxiliary basic instructions are basic
instructions for reading input data from a keyboard, showing output
data on a screen and writing data of a permanent character on a disk.

The maximal internal delay of a \PGLBij\ program concerns the delays
that takes place between successive non-auxiliary basic instructions on
execution of the instruction sequence.
Before we define the maximal internal delay of a \PGLBij\ program, we
define the execution traces of a \PGLBij\ program.

Let $u_1 \conc \ldots \conc u_k$ be a \PGLBij\ program.
Then $\traces(\rho,j,u_1 \conc \ldots \conc u_k)$, where
$\funct{\rho}{[1,\maxr]}{[1,\maxn]}$ and $j \in \Nat$, is the set of
all finite sequences over the set of all primitive instructions of
\PGLBij\ that may be encountered successively on execution of
$u_1 \conc \ldots \conc u_k$ if execution starts with $u_j$ with the
registers used for indirect jumps set according to $\rho$.

Formally, for each \PGLBij\ program $u_1 \conc \ldots \conc u_k$, the
different sets $\traces(\rho,j,u_1 \conc \ldots \conc u_k)$ are defined
by simultaneous induction on the structure of the finite sequences over
the set of all primitive instructions of \PGLBij\ by the following
clauses:
\begin{enumerate}
\item
$\emptyseq \in \traces(\rho,j,u_1 \conc \ldots \conc u_k)$;
\item
if $u_j \equiv a$ or $u_j \equiv \ptst{a}$ or
$u_j \equiv \ntst{a}$, and
$\sigma \in \traces(\rho,j+1,u_1 \conc \ldots \conc u_k)$,
then $u_j \sigma \in \traces(\rho,j,u_1 \conc \ldots \conc u_k)$;
\item
if $u_j \equiv \ptst{a}$ or $u_j \equiv \ntst{a}$, and
$\sigma \in \traces(\rho,j+2,u_1 \conc \ldots \conc u_k)$,
then $u_j \sigma \in \traces(\rho,j,u_1 \conc \ldots \conc u_k)$;
\item
if $u_j \equiv \fjmp{l}$ and
$\sigma \in \traces(\rho,j+l,u_1 \conc \ldots \conc u_k)$,
then
$u_j \sigma \in \traces(\rho,j,u_1 \conc \ldots \conc u_k)$;
\item
if $u_j \equiv \bjmp{l}$ and
$\sigma \in \traces(\rho,j \monus l,u_1 \conc \ldots \conc u_k)$,
then
$u_j \sigma \in \traces(\rho,j,u_1 \conc \ldots \conc u_k)$;%
\footnote
{As usual, we write $i \monus j$ for the monus of $i$ and $j$, i.e.\
 $i \monus j = i - j$ if $i \geq j$ and $i \monus j = 0$ otherwise.
}
\item
if $u_j \equiv \setr{i}{n}$ and
$\sigma \in \traces(\rho',j+1,u_1 \conc \ldots \conc u_k)$,
then
$u_j \sigma \in \traces(\rho,j,u_1 \conc \ldots \conc u_k)$,
where $\rho'$ is defined by $\rho'(i') = \rho(i')$ if $i' \neq i$ and
$\rho'(i) = n$;
\item
if $u_j \equiv \ifjmp{i}$ and
$\sigma \in \traces(\rho,j+\rho(i),u_1 \conc \ldots \conc u_k)$,
then
$u_j \sigma \in
 \traces(\rho,j,u_1 \conc \ldots \conc u_k)$;
\item
if $u_j \equiv \ibjmp{i}$ and
$\sigma \in
 \traces(\rho,j \monus \rho(i),u_1 \conc \ldots \conc u_k)$,
then
$u_j \sigma \in
 \traces(\rho,j,u_1 \conc \ldots \conc u_k)$;
\item
if $u_j \equiv \halt$,
then $u_j \in \traces(\rho,j,u_1 \conc \ldots \conc u_k)$.
\end{enumerate}

For example,
\begin{ldispl}
\traces(\rho_0,1, \ptst{a} \conc \fjmp{3} \conc \setr{1}{3} \conc
                  \fjmp{2} \conc \setr{1}{1} \conc \ifjmp{1} \conc
                  b \conc \fjmp{2} \conc c \conc \halt)\;,
\end{ldispl}%
where $\rho_0$ is defined by $\rho_0(i) = 0$ for all $i \in [1,\maxr]$,
contains
\begin{ldispl}
\ptst{a}\;\; \fjmp{3}\;\; \setr{1}{1}\;\; \ifjmp{1}\;\;
b\;\; \fjmp{2}\;\; \halt\;,
\\
\ptst{a}\;\; \setr{1}{3}\;\; \fjmp{2}\;\; \ifjmp{1}\;\;
c\;\; \halt\;,
\end{ldispl}%
and all prefixes of these two sequences, including the empty sequence.

The set of \emph{execution traces} of a \PGLBij\ program $P$, written
\sloppy
$\traces(P)$, is $\traces(\rho_0,1,P)$, where $\rho_0$ is defined
by $\rho_0(i) = 0$ for all $i \in [1,\maxr]$.

The \emph{maximal internal delay} of an \PGLBij\ program $P$, written
$\MID(P)$, is the largest $n \in \Nat$ for which there exists an
execution trace $u_1 \ldots u_k \in \traces(P)$ and $i_1,i_2 \in [1,k]$
with $i_1 \leq i_2$ such that $\ID(u_j) \neq 0$ for all
$j \in [i_1,i_2]$ and $\sum_{j \in [i_1,i_2]} \ID(u_j) = n$, where
$\ID(u)$ is defined as follows:
\begin{ldispl}
\begin{seqncol}
\ID(a) = 0        & \mif a \notin \Aux\;, \\
\ID(a) = 1        & \mif a \in \Aux\;, \\
\ID(\ptst{a}) = 0 & \mif a \notin \Aux\;, \\
\ID(\ptst{a}) = 1 & \mif a \in \Aux\;, \\
\ID(\ntst{a}) = 0 & \mif a \notin \Aux\;, \\
\ID(\ntst{a}) = 1 & \mif a \in \Aux\;,
\end{seqncol}
\qquad\qquad
\begin{eqncol}
\ID(\fjmp{l}) = 1\;, \\
\ID(\bjmp{l}) = 1\;, \\
\ID(\setr{i}{n}) = 1\;, \\
\ID(\ifjmp{i}) = 2\;, \\
\ID(\ibjmp{i}) = 2\;, \\
\ID(\halt) = 0\;.
\end{eqncol}
\end{ldispl}

Suppose that in the example given above $a$, $b$ and $c$ are
non-auxiliary basic instructions.
Then
\begin{ldispl}
\MID(\ptst{a} \conc \fjmp{3} \conc \setr{1}{3} \conc
     \fjmp{2} \conc \setr{1}{1} \conc \ifjmp{1} \conc
     b \conc \fjmp{2} \conc c \conc \halt) = 4\;.
\end{ldispl}%
This delay takes place between the execution of $a$ and the execution
of $b$ or $c$.

In~\cite{BZ08a}, an extension of basic thread algebra is proposed which
allows for internal delays to be described and analysed.
We could give a formal description of the behaviour of \PGLBij\
programs on execution, internal delays included, using this extension
of basic thread algebra.
Founded on such a formal description, we could explain why the notion
of maximal internal delay of a \PGLBij\ program defined above is the
right one.

$\MID(P)$ can be looked upon as the number of time units that elapses
during the largest possible internal delay of $P$.
Because the time that it takes to execute one basic instruction is
taken for the time unit in the definition of $\MID(P)$, it can
alternatively be looked upon as the number of basic instructions that
can be executed during the largest possible internal delay of $P$.
As usual, the time that it takes to execute one basic instruction is
called a \emph{step}.

\section{Indirect Jumps and Instruction Sequence Performance}
\label{sect-PGLBij-to-PGLB}

In this section, we show that indirect jump instructions are needed for
good instruction sequence performance.

It is assumed that $\bool{1},\bool{2},\ldots \in \Foci$ and
$\setb{\True},\setb{\False},\getb \in \Meth$.
The foci $\bool{1}$, $\bool{2}$, \ldots serve as names of services that
act as Boolean cells.
The methods $\setb{\True}$, $\setb{\False}$, and $\getb$ are accepted by
services that act as Boolean cells and their processing by such a
service goes as follows:
\begin{itemize}
\item
$\setb{\True}$\,:
the contents of the Boolean cell is set to $\True$, and $\True$ is
produced;
\item
$\setb{\False}$\,:
the contents of the Boolean cell is set to $\False$, and $\False$ is
produced;
\item
$\getb$\,:
nothing is changed, but the contents of the Boolean cell is produced.
\end{itemize}
On execution of a \PGLBij\ program that interacts with a Boolean cell,
the content of the Boolean cell may change at any time by external
causes, i.e.\ by causes that do not originate from the executed
program.
The use operators introduced in~\cite{BP02a} permit encapsulation of
services, as a result of which changes by external causes are excluded.
On purpose, we do not apply these use operators here: the Boolean cell
with which we are concerned serves as a means for a program to interact
with the environment on execution.

Below, we will write $\Conc{i = 1}{n} P'_i$, where $P'_1,\ldots,P'_n$
are \PGLBij\ programs, for the \PGLBij\ program
$P'_1 \conc \ldots \conc P'_n$.

For each $k \in \Nat$, let $P_k$ be the following \PGLBij\ program:
\begin{ldispl}
\Conc{i = 1}{2^k}
 (\ntst{\bool{1}.\getb} \conc \fjmp{3} \conc
  \setr{1}{2{\mul}i{+}1} \conc \fjmp{(2^k{-}i){\mul}4{+}2})
 \conc \halt \conc {} \\
\Conc{i = 1}{2^k}
 (\ntst{\bool{1}.\getb} \conc \fjmp{3} \conc
  \setr{2}{2{\mul}i{+}1} \conc \fjmp{(2^k{-}i){\mul}4{+}2})
 \conc \halt \conc {} \\
\ifjmp{1} \conc
\Conc{i = 1}{2^k} (a_i \conc \fjmp{(2^k{-}i){\mul}2{+}1}) \conc
\ifjmp{2} \conc
\Conc{i = 1}{2^k} (a'_i \conc \halt)\;,
\end{ldispl}%
where $a_1,\ldots,a_{2^k}$ and $a'_1,\ldots,a'_{2^k}$ are mutually
different basic instructions.

First, $P_k$ repeatedly tests the Boolean cell $\bool{1}$.
If $\True$ is not returned for $2^k$ tests, $P_k$ terminates.
Otherwise, in case it takes $i$ tests until $\True$ is returned, the
content of register $1$ is set to $2 \mul i + 1$.
If $P_k$ has not yet terminated, it once again repeatedly tests
the Boolean cell $\bool{1}$.
If $\True$ is not returned for $2^k$ tests, $P_k$ terminates.
Otherwise, in case it takes $j$ tests until $\True$ is returned, the
content of register $2$ is set to $2 \mul j + 1$.
If $P_k$ has not yet terminated, it performs $a_i$ after an indirect
jump and following this $a'_j$ after another indirect jump.
After that, $P_k$ terminates.
The length of $P_k$ is $12 \mul 2^k + 4$ primitive instructions and the
maximal internal delay of $P_k$ is $4$ steps.

The \PGLBij\ programs $P_1,P_2,\ldots$ defined above will be used in
the proof of the result concerning the elimination of indirect jump
instructions stated below.

In the proof concerned, we will further make use of the notion of a
reachable occurrence of a basic instruction in a \PGLBij\ program.
We assume that this notion is intuitively sufficiently clear to grasp
the proof.
Given a formal description of the behaviour of \PGLBij\ programs in the
style used for \PGLCij\ programs in~\cite{BM07e}, reachability can
easily be defined in the way followed in Section~3.3 of~\cite{Sch10a}.

Below, we will write $\ell(P)$, where $P$ is a \PGLBij\ program, for the
length of $P$, i.e.\ its number of primitive instructions.

A mapping $\pproj$ from the set of all \PGLBij\ programs to the set of
all \PGLB\ programs has a
\emph{linear upper bound on the increase in maximal internal delay}
if for some $c',c'' \in \Nat$, for all \PGLBij\ programs $P$,
$\MID(\pproj(P)) \leq c' \mul \MID(P) + c''$.
\linebreak[2]
A mapping $\pproj$ from a subset $\mathcal{P}$ of the set of all
\PGLBij\ programs to the set of all \PGLB\ programs has a
\emph{quadratic lower bound on the increase in length}
if\linebreak[2] for some $c',c'' \in \Nat$ with $c' \neq 0$, for all
$P \in \mathcal{P}$, $\ell(\pproj(P)) \geq c' \mul \ell(P)^2 + c''$.

It follows easily from the behaviour-preserving mapping from the set of
all \PGLCij\ programs to the set of all \PGLC\ programs given
in~\cite{BM07e} and an idea used in Section~5.1 of~\cite{BP02a}
that there exists a behaviour-preserving mapping from the set of all
\PGLBij\ programs to the set of all \PGLB\ programs with a linear upper
bound on the increase in maximal internal delay.%
\footnote
{Here behaviour-preserving means preserving the behaviour referred to
 at the end of Section~\ref{sect-PGLBij}, which behaviour does not
 include internal delays.}

\begin{theorem}
\label{theorem-ij-elimination}
Suppose $\pproj$ is a behaviour-preserving mapping from the set of all
\PGLBij\ programs to the set of all \PGLB\ programs with a linear upper
bound on the increase in maximal internal delay.
Moreover, suppose that the number of basic instructions is not bounded.
Then there exists a set $\mathcal{P}$ of \PGLBij\ programs such that
the restriction of $\pproj$ to $\mathcal{P}$ has a
quadratic lower bound on its increase in length.
\end{theorem}
\begin{proof}
For each $k \in \Nat$, let $P_k$ be defined as above.
We show that the restriction of $\pproj$ to $\set{P_1, P_2,\ldots}$ has
a quadratic lower bound on its increase in length.
Take an arbitrary $k \in \Nat$.
Because $\pproj$ has a linear upper bound on the increase in maximal
internal delay, we have
$\MID(\pproj(P_k)) \leq c' \mul \MID(P_k) + c'' = c' \mul 4 + c''$
for some $c',c'' \in \Nat$.
Let $c = c' \mul 4 + c''$.
Suppose that $k$ is much greater than $c$.
This supposition requires that the number of basic instructions is not
bounded.
Because $\pproj$ is a behaviour-preserving mapping and there is a
possibility that $\pproj(P_k)$ contains auxiliary basic instructions,
there are at most $2^c$ different occurrences of basic instructions in
$\pproj(P_k)$ reachable in $c$ steps from any occurrence of a basic
instruction in $\pproj(P_k)$.
Let $i \in [1,2^k]$.
Then, because $\pproj$ is a behaviour-preserving mapping, for each
$j \in [1,2^k]$, some occurrence of $a'_j$ in $\pproj(P_k)$ is
reachable in $c$ steps from each occurrence of $a_i$ in $\pproj(P_k)$.
Thus, there are at least $2^k$ different occurrences of basic
instructions to reach in $c$ steps from one occurrence of $a_i$,
although at most $2^c$ occurrences of basic instructions are reachable
in $c$ steps.
Therefore, there must be at least $2^k / 2^c = 2^{k-c}$ different
occurrences of $a_i$ in $\pproj(P_k)$.
Consequently,
$\ell(\pproj(P_k)) \geq
 2^k \mul 2^{k - c} = 2^{2 \mul k - c}$.
Moreover, $\ell(P_k) = 12 \mul 2^k + 4$.
Hence, the restriction of $\pproj$ to $\set{P_1, P_2,\ldots}$ has
a quadratic lower bound on its increase in length.
\qed
\end{proof}

We conclude from Theorem~\ref{theorem-ij-elimination} that we are faced
with super-linear increases of maximal internal delays if we strive for
acceptable increases of program lengths on elimination of indirect jump
instructions.
In other words, indirect jump instructions are needed for good
instruction sequence performance.

\section{Projectionism}
\label{sect-projectionism}

Semantically, we can eliminate indirect jump instructions by means of a
behaviour-preserving mapping from \PGLBij\ programs to \PGLB\ programs,
but we meet here two challenges of a point of view concerning
programming language semantics of which such behaviour-preserving
mappings form part.

In~\cite{BM09f}, the name projectionism is coined for the point of view
referred to above and its main challenges are discussed.
To put it briefly, projectionism is the point of view that:
\begin{itemize}
\item
any program $P$ first and for all represents a single-pass instruction
sequence as considered in program algebra;
\item
this single-pass instruction sequence, found by behaviour-preserving
mappings called projections, represents in a natural and preferred way
what is supposed to take place on execution of $P$;
\item
program algebra provides the preferred notation for single-pass
instruction sequences.
\end{itemize}
Among the main challenges of projectionism identified in~\cite{BM09f}
are explosion of size, degradation of performance, complexity of
description, and aesthetic degradation.

Projectionism is illustrated in~\cite{BL02a} by means of a hierarchy of
program notations rooted in program algebra.
For each of the program notations that appear in the hierarchy, except
program algebra, a mapping is given by means of which each program
from that program notation is translated into a program from the first
program notation lower in the hierarchy that produces the same behaviour
on execution.
Thus, the single-pass instruction sequence represented by a program in
any program notation from the hierarchy can be found.
The mappings concerned are examples of the projections referred to
above in the outline of projectionism.

The behaviour-preserving mapping from
Theorem~\ref{theorem-ij-elimination} is a projection in the same sense.
Theorem~\ref{theorem-ij-elimination} concerns two of the
above-mentioned challenges of projectionism, for it expresses that in
the case of a projection from \PGLBij\ programs to \PGLB\ programs
there is a trade-off between explosion of size and degradation of
performance.

\section{Conclusions}
\label{sect-conclusions}

In the literature on computer architecture, hardly anything can be
found that contributes to a sound understanding of the effects of the
presence of common kinds of instructions in the instruction set of a
computer on points such as instruction sequence size and instruction
sequence performance.
As a first step towards such an understanding, we have shown that, in
the case where the number of instructions is not bounded, there exist
instruction sequences with direct and indirect jump instructions from
which elimination of indirect jump instructions is possible without a
super-linear increase of their maximal internal delay on execution only
if their lengths increase super-linearly on elimination of indirect
jump instructions.
It is an open problem whether this result goes through in the case where
the number of instructions is bounded.

Instruction sequences with direct jump instructions, indirect jump
instructions and register transfer instructions are as
expressive as instruction sequences with direct jump instructions and
indirect jump instructions without register transfer instructions.
We surmise that a mapping that eliminates the register transfer
instructions yields a result comparable to
Theorem~\ref{theorem-ij-elimination}.
However, we have not yet been able to provide a proof for that case.
On the face of it, a proof for that case is much more difficult than
the proof for the case treated in this paper.

For completeness, we mention that, in the line of research to which the
work presented in this paper belongs, work that is mainly concerned with
direct jump instructions includes the work presented in~\cite{BM08h}.

\bibliographystyle{splncs03}
\bibliography{IS}

\end{document}